# Stability Analysis of Non-Newtonian Rimming Flow

Sergei Fomin[1]*, Ravi Shankar[1], Peter Haine[2]
[1] Department of Mathematics and Statistics, CSU Chico, Chico, CA 95929
[2] Department of Mathematics, Massachusetts Institute of Technology, Cambridge, MA 02139
*Corresponding author; e-mail: sfomin@csuchico.edu, tel.:530-898-5274, fax:530-898-4363

**Abstract**
The rimming flow of a viscoelastic thin film inside a rotating horizontal cylinder is studied theoretically. Attention is given to the onset of non-Newtonian free-surface instability in creeping flow. This non-inertial instability has been observed in experiments, but current theoretical models of Newtonian fluids can neither describe its origin nor explain its onset. This study examines two models of non-Newtonian fluids to see if the experimentally observed instability can be predicted analytically. The non-Newtonian viscosity and elastic properties of the fluid are described by the Generalized Newtonian Fluid (GNF) and Second Order Viscoelastic Fluid (SOVF) constitutive models, respectively. With linear stability analysis, it is found that, analogously to the Newtonian fluid, rimming flow of viscous non-Newtonian fluids (modeled by GNF) is neutrally stable. However, the viscoelastic properties of the fluid (modeled by SOVF) are found to contribute to the flow destabilization. The instability is shown to increase as the cylinder rotation rate is lowered, from which the fluid accumulates in a pool on the rising wall. Viscoelastic effects coupled with this pooling cause the fluid's angular stretching, which is suggested to be responsible for this onset of instability.

## 1. Introduction

The problem of rimming flow has been investigated for many years because of its applications in industry. In most applications, a uniform, smooth film coating is desired [1]. However, experimentally, it was shown that rimming flow is characterized by wide variety of uneven, bulging steady-state film distributions and instabilities [2-7]. Depending on various physical parameters including cylinder rotation rate and cylinder filling fraction, this desirable smooth flow regime may or may not exist. Recently, Seiden and Thomas [8] and Seiden and Steinberg [9] experimentally studied a non-inertial instability unique to viscoelastic rimming flows. Even small concentrations of polymer resulted in free surface plume formation on the rising wall pools, giving rise to complex coagulation dynamics. It is of both practical and theoretical interest to describe such instability analytically.

Previous theoretical studies of rimming flow, however, have largely considered Newtonian fluids. Moffatt [10] first derived the value of the maximum amount of fluid a rotating cylinder can sustain. For masses above this critical value, gravitational forces overcome the cylinder's rotational drag and cause a fluid puddle to accumulate on the rising wall of the cylinder. In the lubrication approximation, O'Brien [11] showed that the position of the puddle on the cylinder wall can be represented by shock solutions. It was later shown that these shock solutions are stable [12-17]. However, these "pooling" solutions exhibit uneven bulges, and can be undesirable for coating applications.

Rimming flow does exhibit smooth free surfaces for subcritical loads. The subcritical regime is characterized by small cylinder filling fraction and fast rotation rate. While rather smooth and uniform, the subcritical film has shown instability in experimental investigations. O'Brien [18] first considered the stability of the subcritical regime for Newtonian fluids. In a linear stability analysis, he showed that for this class of fluids the uniform subcritical solutions are neutrally stable. This linear stability analysis was extended in [19-24] by including higher order effects normally ignored in the lubrication approximation. Pressure differences at the top of the cylinder proved destabilizing, but adding surface tension stabilized the solution [25]. Inertia, however, demonstrated a significant destabilizing effect [23,26].

Although previous studies take many complicated effects into account, they do not consider non-Newtonian rheological effects when studying steady-state stability. Coating industries use polymers (e.g. such as polyethylene) that exhibit viscoelastic rheology that greatly deviates from a Newtonian behavior [27]. Polymers exhibit Newtonian rheology for small strains, but transition to shear-thinning for larger shear rates. They also exhibit much larger normal stresses than Newtonian fluids, so elongation and tension effects become significant [27]. To completely describe this important manufacturing process and the new experimentally detected instabilities seen at low Reynolds number [8-9], the effects of these non-Newtonian properties need to be characterized.

However, the non-Newtonian rimming flow, in general, and its stability, in particular, has not been extensively studied. Fomin *et al.* [28-29] proved that shear-thinning fluids described by the power-law, Ellis, and Carreau models lowered the maximal supportable load of the cylinder. Because shear-thinning inhibits the shear-force of the viscous drag of the cylinder, higher rotation rates are required to offset this gravitational-viscous imbalance. Through numerical investigation, Rajagopalan *et al.* [30] demonstrated that viscoelasticity raises the maximal supportable load.

In our study, the linear stability results obtained for Newtonian fluids are extended to non-Newtonian fluids. The effects of non-Newtonian shear thinning and of elastic normal stresses on the stability of subcritical steady-state rimming flow are studied with a restriction of small Deborah number. To solve the evolution equation governing time-dependent film thickness, the film's free surface is expanded as a normal mode perturbation of the steady-state and the resulting eigenvalue problem is solved. It is shown that, within the lubrication approximation, shear-thinning films are neutrally stable. It is further shown that viscoelasticity destabilizes the film's steady-state. A mechanism is proposed to explain the onset of the experimentally observed instability [8, 9].

## 2. System Model and Scale Analysis

Figure 1 contains a schematic of rimming flow. A horizontal cylinder of radius $r_0$ is rotating in a counterclockwise direction $\theta$ with constant angular velocity $\Omega$ (discussion about a more realistic geometry with tilted cylinders can be found in [31]). A thin liquid film of thickness $h^*(\theta, t)$ moves along the inner cylinder wall due to the gravity and the cylinder's rotational drag force. A cylindrical system of coordinates $(r, \theta, z)$ is used such that the $z$-axis coincides with the axis of the cylinder. The rest of the cylinder is modeled as being filled with rarefied gas of uniform pressure and negligible viscous traction at the liquid-gas interface. It is assumed the cylinder is sufficiently long such that the flow is two-dimensional. Three-dimensional effects have been discussed in some recent publications [32].

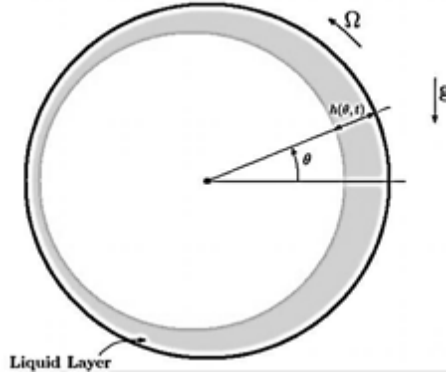

Fig. 1 A simple sketch of the rimming flow system.

The governing equations for incompressible creeping flow are presented:
$$\nabla^* \cdot \boldsymbol{v}^* = 0 \qquad (1)$$
$$\rho \boldsymbol{g} - \nabla^* p^* + \nabla^* \cdot \boldsymbol{\tau}^* = \boldsymbol{0} \qquad (2)$$
where vectors and tensors are denoted in boldface and dimensional variables with asterisks, $\nabla^*$ is the gradient, $\boldsymbol{v}^*$ is the fluid velocity vector with radial and angular components $v_r^*$ and $v_\theta^*$, $p^*$ is the pressure, $\boldsymbol{g}$ is the gravity acceleration vector, and $\boldsymbol{\tau}^*$ is the stress tensor deviator. Expressions and scaling laws for $\boldsymbol{\tau}^*$ are specific to the constitutive model being used. Because two non-Newtonian constitutive models are being considered, $\boldsymbol{\tau}^*$ is addressed later in more detail for each case. The equations (1)-(2) in polar coordinates can be presented as follows:
$$\partial_{r^*}(r^* v_r^*) + \partial_\theta v_\theta^* = 0 \qquad (3)$$
$$-\rho g \sin\theta - \partial_{r^*} p^* + \partial_{r^*} \tau_{rr}^* + (\tau_{rr}^* - \tau_{\theta\theta}^*)/r^* + \partial_\theta \tau_{r\theta}^*/r^* = 0 \qquad (4)$$
$$-\rho g \cos\theta - \partial_\theta p^*/r^* + \partial_\theta \tau_{\theta\theta}^*/r^* + \partial_{r^*} \tau_{r\theta}^* + \tau_{r\theta}^*/r^* = 0 \qquad (5)$$
where $\partial_{r^*}$ and $\partial_\theta$ denote the partial derivatives with respect to $r^*$ and $\theta$, respectively, and $g$ is the gravitational acceleration. At the film's free surface $r^* = r_0 - h^*(\theta, t^*)$, the normal force balance, the tangential force balance, and the kinematic condition are presented below:
$$-p^* + \boldsymbol{n}^* \cdot \boldsymbol{\tau}^* \cdot \boldsymbol{n}^* = 2\kappa^* \sigma, \quad \boldsymbol{n}^* \cdot \boldsymbol{\tau}^* \cdot \boldsymbol{t}^* = 0, \quad \partial_t^* h^* + v_r^* + v_\theta^* \partial_\theta h^*/r^* = 0 \qquad (6)$$
where $\boldsymbol{n}^*$ is the unit normal vector external to the liquid layer, $\boldsymbol{t}^*$ is the unit tangent vector, $\sigma$ is the surface tension, and $\kappa^*$ is the mean curvature of the free surface. The curvature is calculated from $2\kappa^* = \nabla^* \cdot \boldsymbol{n}^*$. The vectors tangent and the normal to the free surface are given by the following equations:
$$\boldsymbol{t}^* = (-\partial_\theta h^* \boldsymbol{e}_r + r^* \boldsymbol{e}_\theta)/\left((\partial_\theta h^*)^2 + r^{*2}\right)^{1/2}, \quad \boldsymbol{n}^* = -(\boldsymbol{e}_r + \partial_\theta h^*/r^* \boldsymbol{e}_\theta)/(1 + (\partial_\theta h^*/r^*)^2)^{1/2}$$
where $\boldsymbol{e}_r$ and $\boldsymbol{e}_\theta$ are the radial and angular basis vectors. On the wall of the cylinder $r^* = r_0$, the no-penetration and no-slip conditions are applied such that $\boldsymbol{v}^* = (v_r^*, v_\theta^*) = (0, \Omega r_0)$.

To proceed further with the analysis, a constitutive model for $\boldsymbol{\tau}^*$ must be specified. The Generalized Newtonian Fluid model is considered first and followed by a discussion of Second Order Viscoelastic Fluids.

## 3. Generalized Newtonian Fluids
Generalized Newtonian Fluids (GNFs or purely viscous fluids) admit a straightforward constitutive model that well describes polymer shear-thinning at large shear rates (see Chapter 4 in [27] for a thorough treatment of this rheological model). This simple constitutive law is given by $\boldsymbol{\tau}^* = 2\mu^* \boldsymbol{\gamma}^*$, where $\boldsymbol{\tau}^*$ is the stress tensor deviator, $\boldsymbol{\gamma}^*$ is the rate of deformation tensor (see the Appendix for tensor components), and $\mu^* = \mu^*(\dot{\gamma}^*)$ is the dynamic viscosity dependent on the shear rate $\dot{\gamma}^* = \sqrt{2\mathrm{tr}(\boldsymbol{\gamma}^* \cdot \boldsymbol{\gamma}^*)}$. The power-law fluid, for example, is governed by this viscosity: $\mu^* = k(\dot{\gamma}^*)^{m-1}$, where $m$ is a flow index and $k$ is a material constant. This formulation also encompasses all other models within the category of Generalized Newtonian fluids (e.g. the Ellis and Carreau constitutive equations).

1) *Scale Analysis*

It is convenient for further analysis to convert variables to a nondimensional form. Rimming flow scaling laws for velocity, pressure, time, etc. are well-documented [10-12]. Taking $\Omega^{-1}$ as the fast time scale as proposed in [23] and denoting $\delta = h_0/r_0$ as a small parameter that represents the ratio of the unknown characteristic film thickness $h_0$ to the cylinder radius, the nondimensional variables can be defined as follows:

$$v_\theta^* = (\Omega r_0)v_\theta, \quad v_r^* = (\Omega r_0 \delta)v_R, \quad r^* = r_0(1 - \delta R), \quad p^* = (\rho g r_0)p, \quad t^* = (\Omega^{-1})t \quad (7)$$

where $R$ is a modified radial coordinate. Using the velocity scaling laws given in (7) and the dimensional forms of the deformation tensor components given in the Appendix A, the following rate of deformation tensor and GNF stress scaling is presented for each tensor component (see [29]):

$$\gamma_{rr}^* = (\Omega)\gamma_{RR}, \quad \gamma_{r\theta}^* = (\Omega/\delta)\gamma_{R\theta}, \quad \gamma_{\theta\theta}^* = (\Omega)\gamma_{\theta\theta} \quad (8)$$
$$\tau_{rr}^* = (\mu_0\Omega)\tau_{RR}, \quad \tau_{r\theta}^* = (\mu_0\Omega/\delta)\tau_{R\theta}, \quad \tau_{\theta\theta}^* = (\mu_0\Omega)\tau_{\theta\theta} \quad (9)$$

where $\mu_0$ is a characteristic viscosity (see [29] for details). Assuming the film is very thin, such that terms of $\mathcal{O}(\delta)$ are negligibly small and can be omitted, the nondimensional stress-strain correlations will take the following forms:

$$\tau_{RR} = 2\mu\,\gamma_{RR}, \quad \tau_{R\theta} = 2\mu\,\gamma_{R\theta}, \quad \tau_{\theta\theta} = 2\mu\,\gamma_{\theta\theta} \quad (10)$$
$$\gamma_{RR} = -\partial_R v_R, \quad \gamma_{R\theta} = -\partial_R v_\theta/2, \quad \gamma_{\theta\theta} = \partial_\theta v_\theta \quad (11)$$

with $\mu = \mu(|\dot{\gamma}|)$ and $\dot{\gamma} = \partial_R v_\theta$. Scale analysis of the free surface boundary conditions (6) gives a non-dimensional Bond number $B = \delta^{-2} r_0 \mu_0 \Omega/\sigma$ that reflects the importance of viscous forces over surface tension. Since $B \gg 1$ for rimming flow, surface tension effects can be neglected (i.e. $B^{-1} \to 0$). Neglecting terms of $\mathcal{O}(\delta)$, the nondimensional free surface force balances and the kinematic condition at the free surface given in (6) yield:

$$R = h: \quad p = 0, \quad \tau_{R\theta} = 0 \quad \partial_t h + v_R + v_\theta \partial_\theta h = 0 \quad (12)$$

The no-penetration and no-slip conditions reduce to $\boldsymbol{v} = (v_R, v_\theta) = (0,1)$ at $R = 0$. By balancing the viscous and gravitational forces in the momentum equations (4) and (5), the characteristic thickness of the liquid layer $h_0 = \delta r_0$ is defined in terms of $\delta = \sqrt{\mu_0 \Omega/\rho g r_0}$. The nondimensional mass conservation (3) and momentum equations (4), (5) to $\mathcal{O}(\delta)$ take the following form:

$$\partial_R v_R - \partial_\theta v_\theta = 0 \quad (13)$$
$$\partial_R p = 0 \quad (14)$$
$$\cos\theta + \partial_\theta p + \partial_R \tau_{R\theta} = 0 \quad (15)$$

2) *Solution*

The solutions of equations (13)-(15) subject to free surface boundary conditions (12) are given below:

$$p = 0, \quad \tau_{R\theta} = (h - R)\cos\theta \quad (16)$$

The next step is to equate the relation for $\tau_{R\theta}$ from the formulae (10), (11) with solution (16) and to solve for $v_\theta$. At this stage, Fomin *et al.* [28-29] introduce a monotonically increasing analytic function $G$ given as the inverse of function $F(\dot{\gamma}) = \dot{\gamma}\mu(|\dot{\gamma}|)$ such that $G(x\mu(|x|)) = x$. This ansatz allows expression of the shear rate in explicit form:

$$\dot{\gamma} = -\text{sgn}(\tau_{R\theta})G(|\tau_{R\theta}|) \qquad (17)$$

where the shear rate is given by $\dot{\gamma} = \partial_R v_\theta$. Integrating equation (17) and accounting for the no-slip conditions yields:

$$v_\theta = 1 - \text{sgn}(\tau_{R\theta})\int_0^R G(|\tau_{R\theta}|)dR \qquad (18)$$

Using the mass conservation equation (13) and the kinematic equation at the free surface given in equation (12), the evolution equation governing the film's free surface thickness variation is obtained:

$$\partial_t h(\theta, t) + \partial_\theta \Phi(\theta, t) = 0 \qquad (19)$$

For GNFs, the non-dimensional mass flux $\Phi = \int_0^h v_\theta dR$ through the liquid layer takes the following form [29]:

$$\Phi = h - \text{sgn}(\cos\theta)\int_0^h RG(R|\cos\theta|)dR \qquad (20)$$

3) *Linear Stability Analysis*

We are interested in the stability of the evolution equation (19), where $\Phi$ is defined by the expression (20). In several studies of non-Newtonian effects on thin films of other geometries [33-34], shear-thinning was shown to have marked effects on the systems' instabilities. A stability analysis of the Generalized Newtonian Fluid can show whether or not the surface plume instability [8-9] is in fact due to a similar shearing phenomenon.

We assume $h(\theta, t)$ can be given as the steady-state solution $h_s(\theta)$ of equations (19), (20) perturbed by a small, angularly periodic, time-dependent disturbance of $\mathcal{O}(\epsilon)$:

$$h(\theta, t) = h_s(\theta) + \epsilon\xi(\theta, t) \qquad (21)$$

Substituting the ansatz (21) into (20), the mass flux $\Phi$ becomes:

$$\Phi = h_s + \epsilon\xi - \text{sgn}(\cos\theta)\int_0^{h_s+\epsilon\xi} RG(R|\cos\theta|)dR \qquad (22)$$

The integral term in (22) can be rewritten as such:

$$\int_0^{h_s+\epsilon\xi} RG(R|\cos\theta|)dR = \int_0^{h_s} RG(R|\cos\theta|)dR + \int_{h_s}^{h_s+\epsilon\xi} RG(R|\cos\theta|)dR \qquad (23)$$

Applying the mean-value theorem to the second integral on the right-hand side of (23) yields:

$$\int_{h_s}^{h_s+\epsilon\xi} RG(R|\cos\theta|)dR = \epsilon\xi(h_s + \chi\epsilon\xi)G[(h_s + \chi\epsilon\xi)|\cos\theta|] \qquad (24)$$

where $0 < \chi < 1$. Since $G$ is a continuous function [29],

$$G[(h_s + \chi\epsilon\xi)|\cos\theta|] = G(h_s|\cos\theta|) + p_\epsilon(\theta, t) \qquad (25)$$

where $p_\epsilon(\theta, t) \to 0$ as $\epsilon \to 0$. With equations (23)-(25), and neglecting terms of $o(\epsilon)$ (i.e. linearizing in $\epsilon$), the linearized mass flux $\Phi$ of (22) is given by:

$$\Phi = h_s - \text{sgn}(\cos\theta)\int_0^{h_s} RG(R|\cos\theta|)dR + \epsilon\xi\left(1 - h_s\text{sgn}(\cos\theta)G(h_s|\cos\theta|)\right) \qquad (26)$$

Notice that the leading order terms in (26) cancel since $h_s(\theta)$ satisfies the steady-state form of equations (19), (20), as defined. Combining equations (19) and (26) and taking only leading order terms, the linear evolution equation for the traveling disturbance $\xi(\theta, t)$ is obtained:

$$\partial_t\xi(\theta, t) + \partial_\theta[\alpha(\theta)\xi(\theta, t)] = 0 \qquad (27)$$

Here, $\alpha(\theta) = 1 - h_s(\theta)\text{sgn}(\cos\theta)G(h_s(\theta)|\cos\theta|)$ represents the speed of the disturbance's propagation along the free surface. In [29] it was shown that the steady-state solution $h_s(\theta)$ exists only if $\alpha(\theta) > 0$ for all $\theta$. For a normal-mode analysis (see [35]), we assume that the disturbance $\xi$ is harmonic in time, such that $\xi(\theta, t) = \mathcal{R}e[f(\theta)e^{st}]$. Here, $\mathcal{R}e[]$ denotes the real part, $f(\theta)$ is an unknown complex periodic function, and $s$ is an unknown complex growth factor. Depending on whether the growth rate $\mathcal{R}e[s]$ is greater than, equal to, or less than zero, the steady-state $h_s(\theta)$ can be either unstable, neutrally stable, or asymptotically stable with respect to disturbance $\xi$. Dropping $e^{st}$'s, the evolution equation (27) reduces to the following eigenvalue problem for $f(\theta)$ and $s$:

$$-sf(\theta) = \partial_\theta[\alpha(\theta)f(\theta)] \tag{28}$$

Since $\alpha(\theta) \neq 0$ according to [29], equation (28) has nonzero solution $(\theta)$. Solution of this equation is straightforward:

$$f(\theta) = A\alpha(\theta)^{-1}\exp(-s\int_0^\theta \alpha(\phi)^{-1}d\phi) \tag{29}$$

where $A$ is an integration constant. This is analogous to what O'Brien obtains when specifically considering Newtonian fluids [18]. Applying to $f(\theta)$ the periodic boundary condition $f(\theta) = f(\theta + 2\pi)$, and recognizing that $\alpha(\theta)$ is $2\pi$-periodic, it becomes clear that $1 = \exp(-s\int_0^{2\pi}\alpha(\phi)^{-1}d\phi)$. Since $\alpha(\theta) > 0$ for all $\theta$, the integral term is nonzero, such that we must have the following condition for $s$:

$$s = 2\pi in/\int_0^{2\pi}\alpha(\phi)^{-1}d\phi \tag{30}$$

Since $\alpha$ is real, $s$ must be purely imaginary ($\mathcal{R}e[s] = 0$) for all integers (wavenumbers) $n$, which is related to the disturbance's spatial frequency. Therefore, since $\mathcal{R}e[s] = 0$, it can be said that steady-state solutions for Generalized Newtonian Fluid rimming flows are neutrally stable to small perturbations. This result holds for many constitutive models, including the simple power-law model as well as the more sophisticated and realistic Ellis and Carreau models. Interestingly, the power-law model seems to describe free surface perturbations of rimming flow steady-states just as well as the Carreau model.

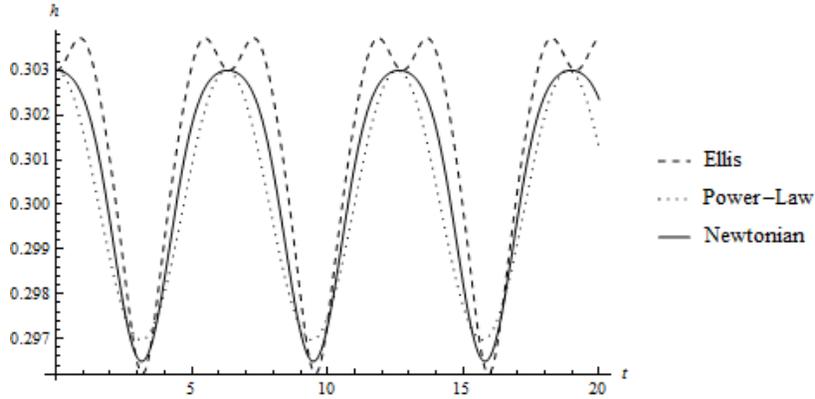

Fig. 2 Numerical solution of equation (19) and (20) for Newtonian, power-law, and Ellis fluids viewed at $\theta = 0$ for $h_0 = 0.3, m = 0.3, Wi = 3$, and $\epsilon = 0.01$.

To illustrate these neutral stability results numerically, the power-law and Ellis models are considered. For the power-law model, we have $\mu(\dot{\gamma}) = |\dot{\gamma}|^{m-1}$ and $G(x) = |x|^{1/m}$, where m is a flow index [29]. For $m = 1$, the model is Newtonian, and for $0 < m < 1$, the model describes shear-thinning that occurs for polymers at large shear rates [27]. The Ellis model includes a more realistic and gradual transition from Newtonian to shear-thinning behavior as the shear-rate is increased [27]. For the Ellis model, we have $\dot{\gamma} = -\tau_{\theta R}\left(1 + (Wi|\tau_{\theta R}|)^{\frac{1}{n}-1}\right)$ and $G(x) = x\left(1 + (Wi|x|)^{\frac{1}{m}-1}\right)$, where $0 < m < 1/3$ is a flow index and $Wi = \lambda\Omega/\delta$ is a non-dimensional shear-thinning number for time constant $\lambda$ [29]. For $Wi = 0$, the model is Newtonian. Equations (19) and (20) are solved with the Newtonian, power-law and Ellis models. The initial condition is given for each model as the corresponding steady-state distribution perturbed by a small periodic disturbance, specifically $h(\theta, 0) = h_s(\theta) + \epsilon h_0 \cos\theta$, where $\epsilon \ll 1$ is a small parameter, and $h_0 = h_s(0)$. Each solution in Fig. 2 appears periodic and has bounded, unchanging amplitude given by the initial perturbation. These computations illustrate the analytically proven neutral stability of the viscous non-Newtonian steady-state rimming flow. In other words, neither the Newtonian approximation nor the shear-thinning corrections can describe the surface plume instability. This is consistent with experimental observations [8, 9].

## 4. Second Order Viscoelastic Fluids

In this section, the fluid's normal-stress properties are considered. To qualitatively investigate the influence of weak viscoelasticity on rimming flow and its stability, its behavior can be modeled with a simple quasilinear constitutive equation:

$$\boldsymbol{\tau}^* = 2\mu_0 \boldsymbol{\gamma}^* - 2\mu_0 \lambda_0 \hat{\boldsymbol{\gamma}}^* \tag{31}$$
$$\hat{\boldsymbol{\gamma}}^* = \partial_{t^*} \boldsymbol{\gamma}^* + \boldsymbol{v}^* \cdot \boldsymbol{\nabla}^* \boldsymbol{\gamma}^* - (\boldsymbol{\nabla}^* \boldsymbol{v}^*)^T \cdot \boldsymbol{\gamma}^* + ((\boldsymbol{\nabla}^* \boldsymbol{v}^*)^T \cdot \boldsymbol{\gamma}^*)^T \tag{32}$$

Here, as in the GNF case, $\boldsymbol{\tau}^*$ is the deviator of the viscoelastic stress tensor, $\boldsymbol{\gamma}^*$ is the deformation tensor, $\hat{\boldsymbol{\gamma}}^*$ is the upper-convected time derivative of the deformation tensor $\boldsymbol{\gamma}^*$, $\mu_0$ is again the fluid's characteristic viscosity, and $\lambda_0$ is the fluid's retardation time constant. This model describes fluids with negligible second normal stress coefficients (which is common for many polymers, see Chapter 3 in [27]), a special class of *Second Order Viscoelastic Fluids* (chapter 6 in [27]). The Second Order Viscoelastic Fluid (SOVF) model does not reproduce shear-thinning like the GNF model, so it is restricted to small deformation rates. However, since it has been shown that shear-thinning in GNFs does not affect rimming flow's linear stability, ignoring this aspect of viscoelasticity should simplify computations considerably with little cost to accuracy.

1) *Scale Analysis*

The velocity, pressure, time, and deformation tensor are non-dimensionalized as before (see equations (7) and (8)). Using the dimensional form of $\hat{\boldsymbol{\gamma}}^*$ given in the Appendix, the tensors' components are non-dimensionalized as follows:

$$\hat{\gamma}_{rr}^* = (\Omega)\hat{\gamma}_{RR}, \qquad \hat{\gamma}_{r\theta}^* = (\Omega/\delta)\hat{\gamma}_{R\theta}, \qquad \hat{\gamma}_{\theta\theta}^* = (\Omega/\delta^2)\hat{\gamma}_{\theta\theta} \tag{33}$$
$$\tau_{rr}^* = (\mu_0 \Omega)\tau_{RR}, \qquad \tau_{r\theta}^* = (\mu_0 \Omega/\delta)\tau_{R\theta}, \qquad \tau_{\theta\theta}^* = (\mu_0 \Omega/\delta^2)\tau_{\theta\theta} \tag{34}$$

The new choice of scale for $\tau_{\theta\theta}^*$ comes from the upper-convected derivative $\hat{\gamma}_{\theta\theta}^*$, in which angular elongation terms nonlinearly couple with the shear strain [36-37]. We introduce the nondimensional Deborah number $De = \lambda_0 \Omega$ as a measure of viscoelasticity. SOVF models are restricted to weakly viscoelastic fluids of small $De \ll 1$. Omitting terms of $\mathcal{O}(\delta)$, the nondimensional SOVF tensors components reduce to the following form:

$$\tau_{RR} = 2\gamma_{RR} - 2De\,\hat{\gamma}_{RR}, \qquad \tau_{R\theta} = 2\gamma_{R\theta} - 2De\,\hat{\gamma}_{R\theta}, \qquad \tau_{\theta\theta} = -2De\,\hat{\gamma}_{\theta\theta} \tag{35}$$
$$\hat{\gamma}_{RR} = \partial_t \gamma_{RR} - v_R \partial_R \gamma_{RR} + v_\theta \partial_\theta \gamma_{RR} + 2\gamma_{RR} \partial_R v_R - 2\gamma_{R\theta} \partial_\theta v_R \tag{36}$$
$$\hat{\gamma}_{R\theta} = \partial_t \gamma_{R\theta} - v_R \partial_R \gamma_{R\theta} + v_\theta \partial_\theta \gamma_{R\theta} + \gamma_{RR} \partial_R v_\theta, \qquad \hat{\gamma}_{\theta\theta} = 2\gamma_{R\theta} \partial_R v_\theta \tag{37}$$

The momentum equations (4),(5) become:

$$\partial_R p = 0 \tag{38}$$
$$\cos\theta + \partial_\theta p + \partial_R \tau_{R\theta} - \partial_\theta \tau_{\theta\theta} = 0 \tag{39}$$

The no-penetration and no-slip conditions at $R = 0$:

$$\boldsymbol{v} = (v_R, v_\theta) = (0, 1) \tag{40}$$

with the following normal and transverse force balances on the free surface:

$$R = h: \qquad p = 0, \qquad \tau_{R\theta} + \tau_{\theta\theta} \partial_\theta h = 0 \tag{41}$$

*2) Solution*

Again, the pressure is found to be $p = 0$. The unknowns are expanded as perturbation series in powers of small $De \ll 1$. Each component $\tau_{XY}$ of tensor $\boldsymbol{\tau}$ is expanded as $\tau_{XY} = \tau_{XY0} + De\tau_{XY1} + \mathcal{O}(De^2)$, associating each numerical subscript with the corresponding power of $De$. The components of $\boldsymbol{\gamma}$ and $\hat{\boldsymbol{\gamma}}$ are expanded similarly. Each vector component $v_x$ of $\boldsymbol{v}$ is expanded as $v_x = v_{x0} + Dev_{x1} + \mathcal{O}(De^2)$. Letting terms of $\mathcal{O}(De^2) \to 0$ and collecting terms of each power of $De$ in equations (35)-(41) yields leading order and first order systems of equations. The leading order (Newtonian) solutions are presented to $\mathcal{O}(1)$:

$$\tau_{\theta\theta 0} = 0 \tag{42}$$
$$\tau_{R\theta 0} = (h - R)\cos\theta, \tag{43}$$
$$v_{\theta 0} = 1 + \cos\theta \, (R^2/2 - Rh) \tag{44}$$
$$v_{R0} = -R^3 \sin\theta /6 - R^2 \partial_\theta (h\cos\theta)/2 \tag{45}$$
$$\tau_{RR0} = R^2 \sin\theta + 2R^2 \partial_\theta (h\cos\theta) \tag{46}$$

The first order non-Newtonian corrections of $\mathcal{O}(De)$:

$$\tau_{\theta\theta 1} = 2(h - R)^2 \cos^2\theta \tag{47}$$
$$\tau_{R\theta 1} = -2(h - R)^2 \cos\theta \, (2(R - h)\sin\theta + 3\cos\theta \, \partial_\theta h)/3 \tag{48}$$
$$v_{\theta 1} = R(\partial_\theta + \partial_t)(h\cos\theta) - R^2 \sin\theta /2 + R\partial_\theta\big((2h/3 - R/4)h^2 \cos^2\theta\big) \tag{49}$$

Note that the viscoelastic inclusion yields an $\mathcal{O}(De)$ normal stress (47) in the angular direction not present in the Newtonian model for $\delta \to 0$. Comparing with the shear-stress in (43), this is seen to be a manifestation of nonlinear coupling evident also in dimensional scaling of equation (34).

Employing the velocity expansion $v_\theta = v_{\theta 0} + De \, v_{\theta 1} + \mathcal{O}(De^2)$ and the definition $\Phi = \int_0^h v_\theta dR$, the mass flux for the SOVF model is obtained with accuracy to $\mathcal{O}(De^2)$:

$$\Phi = h - h^3(2\cos\theta - De \sin\theta)/6$$
$$+ De \, \big(h^5 \sin 2\theta /2 + \partial_\theta(h^5 \cos^2\theta) - (\partial_\theta + \partial_t)(h^3 \cos\theta)\big)/6 \tag{50}$$

Following Benilov et al [23], the mass flux (50) can be greatly simplified by evaluating equation (19) to leading order in $De$: $\partial_t h + \partial_\theta h = \partial_\theta (h^3 \cos\theta)/3 + \mathcal{O}(De)$. Substituting this relation into equation (50), an (asymptotically) equivalent but simpler mass flux is obtained to $\mathcal{O}(De^2)$:

$$\Phi = h - \frac{1}{3}h^3(\cos\theta - De\sin\theta) - \frac{1}{6}De \, h^5 \sin 2\theta + \frac{1}{3}De \, h^4 h_\theta \cos^2\theta \tag{51}$$

The last term in the expression (51) comes from balancing the shear stress (48) in momentum equation (39) with the angular extensional force of (47). Interestingly, a similar angular term of $h^6 h_\theta \cos^2\theta$ can be seen in the model of [23] for weak *inertial* corrections. In our study of creeping flow, however, the inertial forces are negligible. That this angular stress term still occurs even in a non-inertial regime illustrates why instability can be seen in both low [8] and high [6] Reynolds number regimes in experiments.

*3) Steady-State Solution*

In the steady-state case (when $\partial_t h = 0$) equation (19) yields $\partial_\theta \Phi_s = 0$, or after integration $\Phi_s = q$, where $q$ is the constant of integration, which physically represents the mass flux through the liquid layer in the steady-state case. So, the steady-state mass flux $\Phi_s$ from (51) is constant. Obviously, parameter $q$ becomes larger for increased cylinder filling fraction or for decreased cylinder rotation rate. A convenient expression for the viscoelastic steady-state flow can be found by expanding the solution of (51) in powers of small $q$:

$$h_s(\theta) \approx q + \frac{1}{3}q^3(\cos\theta - De\sin\theta) + \mathcal{O}(q^5) \tag{52}$$

Figure 3 provides a comparison of the numerical solution of equation (51) to the expression (52). Notice that the $q = 0.1$ case depicts an evenly distributed film. The approximation works well even for larger values of mass flux including $q = 0.5$. This approximation to the steady-state $h_s(\theta)$ will be used in the following stability analysis.

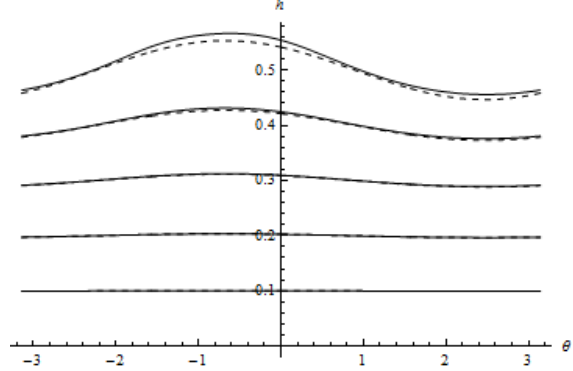

Fig. 3 Viscoelastic steady-state solutions for various values of mass flux $q$ and $De = 0.8$. Solid line is numerical solution of equation (51). Dashed line is analytical expression (52). For increasing heights of curves, $q$ varies as $q = 0.1, 0.2, 0.3, 0.4, 0.5$.

4) *Linear Stability Analysis*

As before, we assume that $h(\theta, t)$, the solution of equations (19), (51), is a small time-dependent, periodic in $\theta$ perturbation of the steady-state solution $h_s(\theta)$ such that $h(\theta, t) = h_s(\theta) + \epsilon \xi(\theta, t)$. After linearizing in $\epsilon$, the evolution equations (19), (51) reduce to:

$$\partial_t \xi = -\partial_\theta \left( \xi(1 - h_s^2(\cos\theta - De\sin\theta) - 7/6\, De h_s^4 \sin 2\theta) \right)$$
$$- De\cos^2\theta\, \partial_\theta^2(h_s^4 \xi)/3 - 2De h_s^4 \xi \cos 2\theta /3 \tag{53}$$

It is interesting to note that the latter equation is a variable-coefficient diffusion equation. The diffusivity dependence on the angular coordinate came from considering normal stress contributions in the SOVF model. The diffusivity $-De h_s^4 \cos^2\theta /3$ is seen to be negative, such that the disturbance $\xi$ is amplified via backward diffusion. So, rather than neutral stability in the purely shearing case explored previously, this viscoelastic steady-state should exhibit *instability* due to addition of normal stress terms.

To quantify this result, equation (53) is again solved with a normal-mode expansion. Because the instability is expected for all steady-states $h_s(\theta)$, stability analysis of (53) can be restricted to small $q \ll 1$. Employing the small $q$ expansion (52) and approximating the coefficients of (53) to $\mathcal{O}(q^6)$, the evolution equation reduces to:

$$\partial_t \xi = -De q^4 \xi_{\theta\theta} \cos^2\theta/3 - 2De q^4 \xi \cos 2\theta/3$$
$$-\partial_\theta \left( \xi(1 - q^2(\cos\theta - De\sin\theta) - q^4/6(4\cos^2\theta + 4De^2\sin^2\theta + 3De\sin 2\theta)) \right) \tag{54}$$

As before, for a normal-mode analysis, we assume that the disturbance $\xi$ can be represented as $\xi(\theta, t) = \mathcal{R}e[f(\theta)e^{st}]$. The evolution equation (54) becomes the following eigenvalue problem for the disturbance's complex spatial profile $f(\theta)$ and complex growth rate $s$:

$$-sf = De q^4 f_{\theta\theta} \cos^2\theta/3 + 2De q^4 f \cos 2\theta/3 + \partial_\theta \left( f(1 - q^2(\cos\theta - De\sin\theta) - q^4/6(4\cos^2\theta + 4De^2\sin^2\theta + 3De\sin 2\theta)) \right) \tag{55}$$

5) *Numerical results*

It is assumed that $f(\theta)$ can be represented by a complex Fourier Series, such that $f = \sum_{n=-\infty}^{+\infty} F_n \exp in\theta$. By computing the action of the differential operator given in (55) on the Fourier series, the differential equation can be discretized into the form of a matrix eigenvalue problem:

$$(\mathbf{G} + s\mathbf{I})\mathbf{f} = 0 \tag{56}$$

where $\mathbf{f}$ is a vector of the Fourier coefficients $F_k$ of $f(\theta)$, and $\mathbf{I}$ is the identity matrix. The coefficient matrix $\mathbf{G}$ can be given as:

$$\mathbf{G} = \begin{pmatrix} \ddots & b_{k-1} & c_{k-1} & d_{k-1} & e_{k-1} & 0 & 0 \\ 0 & a_k & b_k & c_k & d_k & e_k & 0 \\ 0 & 0 & a_{k+1} & b_{k+1} & c_{k+1} & d_{k+1} & \ddots \end{pmatrix} \tag{57}$$

where

$$a_k = \frac{1}{12}kq^4(De(1-k) + 2(De^2 - 1)i)$$

$$b_k = -\frac{1}{2}kq^2(De - i)$$

$$c_k = \frac{1}{6}k(6i - q^4(De\,k + 2(De^2 + 1)i))$$

$$d_k = -\frac{1}{2}kq^2(De + i)$$

$$e_k = -\frac{1}{12}kq^4(De(1+k) - 2(De^2 - 1)i)$$

This type of discretization is not always optimal (see [24, 38]). However, for this case, the results of this method agree quite well with analytical approximations (see the Appendix B). Figures 4 and 5 depict numerically obtained eigenvalues for various values of $q$ and $De$. From each figure, it's clear that the growth rates (i.e. $Re[s]$) are positive for various frequencies $Im[s]$, confirming the existence of instability. From Fig. 4, the effect of rising $De$ is to increase the growth rates for each frequency. The case of very small $De$ shows vanishing real parts, in agreement with results obtained earlier for the purely viscous GNF model. Clearly, the instability is due to non-Newtonian effects, but the comparison of GNF and SOVF results shows that the instability is due to normal-stress interactions rather than shear-thinning.

The instability mechanism can be readily understood from the disturbances' growth-rates' dependence on $q$. From Fig. 5, the instability all but disappears when $q$ is lowered (i.e. when the rotation rate is raised ). In the limit of $q \to 0$, equation (52) evidently gives $h_s(\theta) \to q$, and the film becomes evenly distributed along the wall. As $q$ is raised, the $\cos\theta$ term causes $h_s(\theta)$ to increase near $\theta = 0$, which corresponds to the film pooling on the rising wall due to increasing importance of gravity (see Fig. 3). So, from equation (43), shear-stress at the wall $R = 0$ clearly becomes stronger on the rising side as $q$ is increased. Now, from equation (53), the backwards-diffusion coefficient with $h_s^4 \cos^2\theta$ is clearly responsible for the instability. This term comes from the angular extensional stress of equation (47), which is generated from nonlinear coupling with the shear stress in equation (43). Thus, given an increase in $q$, the resulting increase in shear stress on the rising wall produces a corresponding *quadratic* increase in angular extensional stress at the wall, which induces a *quartic* increase in the backward diffusion term. This level of dependence is consistent with results in Fig. 5 and the analytical approximation of Appendix B.

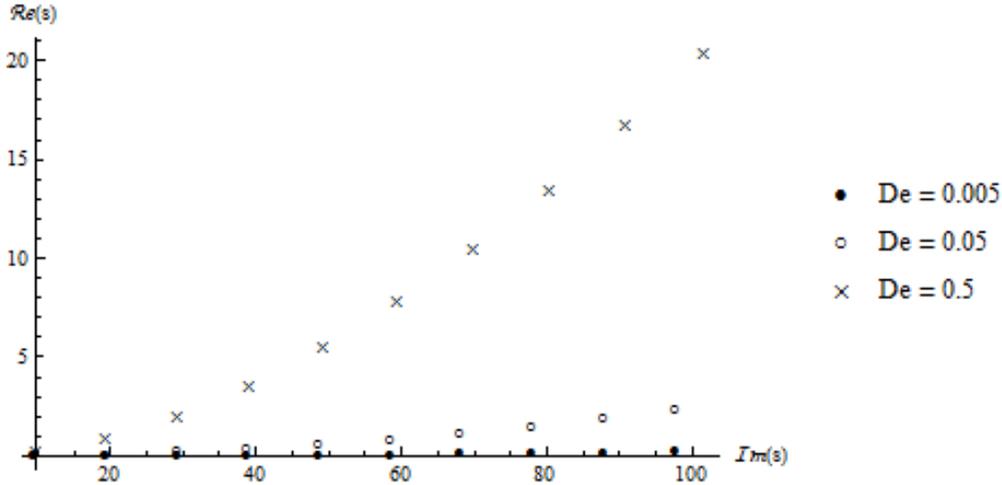

Fig. 4 Numerically obtained eigenvalues for $q = 0.4$. The eigenvalues are computed with wavenumbers $n = -100$ to $0$. The growth rates $\mathcal{R}e(s)$ are parametrically plotted with respect to frequencies $\mathcal{I}m(s)$.

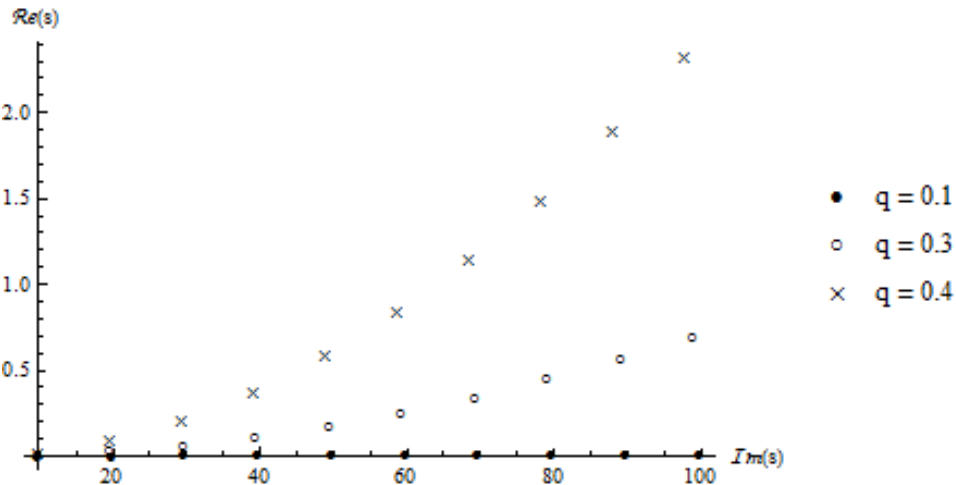

Fig. 5 Numerically obtained eigenvalues for $De = 0.05$. The eigenvalues are computed with wavenumbers $n = -100$ to $0$. The growth rates $\mathcal{R}e(s)$ are parametrically plotted with respect to frequencies $\mathcal{I}m(s)$.

## 5. Conclusions

To summarize, non-Newtonian effects on rimming flow stability have been investigated theoretically to explain a non-inertial instability observed experimentally [8-9] in viscoelastic rimming flow. Different aspects of viscoelasticity, including shear-thinning and enhanced normal stresses, have been simulated with the Generalized Newtonian Fluid and Second Order Viscoelastic Fluid models. Conclusions have been drawn using perturbative approaches in limiting two-dimensional flow regimes, namely the lubrication limit of $\delta \ll 1$ and $Re \ll 1$, and the weakly viscoelastic limit of $De < 1$. For the GNF model of shear-thinning, all purely viscous steady-states were shown to be neutrally stable, as in the Newtonian case, suggesting that the surface plume instability is not caused by viscous non-Newtonian properties of the fluid.

However, using the SOVF model of enhanced normal stresses, the steady-state was instead shown to have instability. The growth rates of small disturbances were shown to vary strongly with increasing $q$, the increasing value of which is associated with pooling of fluid on the cylinder's raising wall as the rotation rate is decreased. For viscoelastic fluids, angular extensional forces are nonlinearly coupled to the shear-stress, as given by the upper-convected time derivative of the strain tensor. So, since the shearing on the rising wall increases with the size of the pooling bulge, the extensional stretching grows quadratically in unison, giving rise to the instability. This is in contrast to Generalized Newtonian films, which exert no extensional stresses for $\delta \ll 1$. These results seem qualitatively consistent with the experimental results of [8, 9], viz. the distorted shape of the free surface pooling (Figure 4a in [9]) and the pronounced extensional flow at the rising wall (Figure 2 in [8]). That instability was so observed in a non-inertial rimming flow can therefore be qualitatively explained by our model for weak viscoelasticity. As a concluding remark, given the strong quartic dependence of the eigenfunction growth rates on $q$ (see Appendix B), it seems unlikely that this instability would be experimentally observable for larger rotation rates (very small $q$), in which the thin film is evenly distributed on the entire wall.

**Acknowledgments**

The authors would like to thank Dr. Marina Chugunova of Claremont Graduate University for her useful comments and advice. This research was funded by the NSF award DMS-1156612.

**Appendix A**

$$\gamma_{rr}^* = \frac{\partial v_r^*}{\partial r^*}, \qquad \gamma_{r\theta}^* = \frac{1}{2}\frac{\partial v_\theta^*}{\partial r^*} + \frac{1}{2r^*}\left(\frac{\partial v_r^*}{\partial \theta} - v_\theta^*\right), \qquad \gamma_{\theta\theta}^* = \frac{1}{r^*}\left(\frac{\partial v_\theta^*}{\partial \theta} + v_r^*\right) \tag{A1}$$

$$\hat{\gamma}_{rr}^* = \frac{\partial \gamma_{rr}^*}{\partial t^*} + v_r^*\left(\frac{\partial \gamma_{rr}^*}{\partial r^*}\right) + \frac{v_\theta^*}{r^*}\frac{\partial \gamma_{rr}^*}{\partial \theta} - 2\frac{v_\theta^*}{r^*}\gamma_{r\theta}^* - 2\frac{\partial v_r^*}{\partial r^*}\gamma_{rr}^* - \frac{2}{r^*}\frac{\partial v_r^*}{\partial \theta}\gamma_{r\theta}^* + \frac{2}{r^*}v_\theta^*\gamma_{r\theta}^* \tag{A2}$$

$$\hat{\gamma}_{r\theta}^* = \frac{\partial \gamma_{r\theta}^*}{\partial t^*} + v_r^*\frac{\partial \gamma_{r\theta}^*}{\partial r^*} + \frac{v_\theta^*}{r^*}\frac{\partial \gamma_{r\theta}^*}{\partial \theta} + \frac{v_\theta^*}{r^*}(\gamma_{rr}^* - \gamma_{\theta\theta}^*) - \frac{\partial v_r^*}{\partial r^*}\gamma_{r\theta}^* - \frac{1}{r^*}\frac{\partial v_r^*}{\partial \theta}\gamma_{\theta\theta}^* + \frac{v_\theta^*}{r^*}\gamma_{\theta\theta}^* - \frac{\partial v_\theta^*}{\partial r^*}\gamma_{rr}^* - \frac{1}{r^*}\frac{\partial v_\theta^*}{\partial \theta}\gamma_{r\theta}^* - \frac{v_r^*}{r^*}\gamma_{r\theta}^* \tag{A3}$$

$$\hat{\gamma}_{\theta\theta}^* = \frac{\partial \gamma_{\theta\theta}^*}{\partial t^*} + v_r^*\frac{\partial \gamma_{\theta\theta}^*}{\partial r^*} + \frac{v_\theta^*}{r^*}\frac{\partial \gamma_{\theta\theta}^*}{\partial \theta} + 2\frac{v_\theta^*}{r^*}\gamma_{r\theta}^* - 2\frac{\partial v_\theta^*}{\partial r^*}\gamma_{r\theta}^* - \frac{2}{r^*}\frac{\partial v_\theta^*}{\partial \theta}\gamma_{\theta\theta}^* - 2\frac{v_r^*}{r^*}\gamma_{\theta\theta}^* \tag{A4}$$

**Appendix B**

The eigenvalue problem (55) is solved in the limit of $q \to 0$ by expanding $s$ and $f(\theta)$ in straightforward power series as $s = \sum_{i=0}^{N} q^{2i} s_i + \mathcal{O}(q^{2N+2})$ and $f(\theta) = \sum_{i=0}^{N} q^{2i} f_i(\theta) + \mathcal{O}(q^{2N+2})$. Terms of each power of $q^2$ are collected, of which the coefficients are set to zero identically, and the associated differential equations are solved for each function $f_i$. Functions $f_i(\theta)$ are required to be periodic such that $f_i(\theta) = f_i(\theta + 2\pi)$. The eigenfunction $f$ and eigenvalue $s$ are found to be:

$$f(\theta) = e^{-in\theta}\left(1 + q^2(\cos\theta - De\sin\theta + in(\cos\theta + De\sin\theta))\right) + \mathcal{O}(q^4) \tag{B1}$$

$$s = De\, q^4 n^2/6 - i\, n(1 - 5q^4(1 + De^2)/6) + \mathcal{O}(q^6) \tag{B2}$$

Where $n = 0, \pm 1, \pm 2 \ldots$ is an integer resulting from application of the periodic boundary condition. To $\mathcal{O}(q^6)$, the real part of $s$ is positive, so disturbances $\xi(\theta, t) = \mathcal{R}e[f(\theta)e^{st}]$ will grow with time, thus confirming that the steady-state given by equation (52) is unstable. As $q \to 0$, the real part is seen to vanish as a quartic, showing that the normal-stress instability disappears in the large rotation rate limit.

From (B1), for $n \sim q^{-2}$, the $\mathcal{O}(q^2)$ correction $f_1(\theta)$ becomes of similar size $\mathcal{O}(1)$ to the leading order approximation $f_0(\theta)$. A suitable WKB rescaling is to be made [24]. The dominant balance is $\lambda = s/De^{-1}$ and $f(\theta) = \exp\left(De^{-1} \int_0^\theta F(\phi) d\phi\right)$. Substituting the ansatz into equation (55) and letting $F(\theta) = F_0(\theta) + \mathcal{O}(De)$, the leading order approximation to $\mathcal{O}(De)$ is found to be:

$$F(\theta) = 1 - \left(3\sqrt{2}(1 - q^2 \cos\theta) - \sqrt{\sum_{j=0}^{4} a_j q^{2j} \cos j\theta}\right) / (2\sqrt{2} q^4 \cos^2\theta), \qquad (B3)$$

where:
$a_0 = 3(6 - q^4(1 + 4\lambda) + q^8)$, $a_1 = -18(2 - q^4)$, $a_2 = 4q^4 - 3(1 + 4\lambda)$, $a_3 = 6$, $a_4 = 1$.

Applying the periodic boundary condition, the far-field eigenvalues $\lambda$ can be computed by numerically solving a transcendental equation: $2\pi i n = De^{-1} \int_0^{2\pi} F(\phi) d\phi$. Figure 6 compares eigenvalues obtained using the numerical solution, the WKB approximation, and the straightforward power series expansion (B2). Clearly, the numerical results agree well with the perturbative solutions.

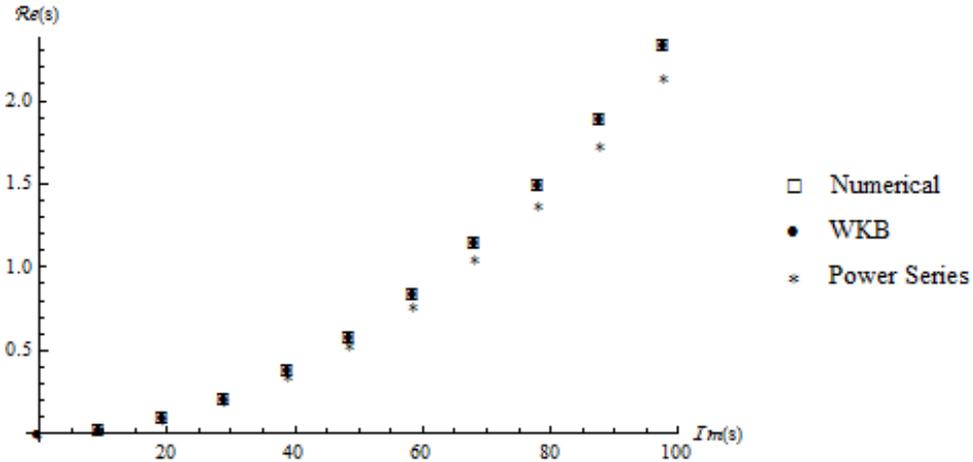

**Fig. 6** Eigenvalues obtained for $q = 0.3$ and $De = 0.05$. The eigenvalues are computed with wavenumbers $n = -100$ to $0$. The disturbances' growth rates $\mathcal{R}e(s)$ are parametrically plotted with respect to frequencies $\mathcal{I}m(s)$.